\documentclass[nofootinbib]{revtex4}
\usepackage{graphicx}
\usepackage{latexsym}
\usepackage{amsthm,amsmath,amssymb}
\usepackage{mathrsfs}
\def\be{\begin{equation}}
\def\ee{\end{equation}}
\def\bea{\begin{eqnarray}}
\def\eea{\end{eqnarray}}

\begin{document}

\hfill  USTC-ICTS/PCFT-20-19

\title{A simple parity violating gravity model without ghost instability}

\author{Mingzhe Li}
%\email{limz@ustc.edu.cn}
\author{Haomin Rao}
%\email{rhm137@mail.ustc.edu.cn}
\author{Dehao Zhao}
%\email{dhzhao@mail.ustc.edu.cn}
\affiliation{Interdisciplinary Center for Theoretical Study, University of Science and Technology of China, Hefei, Anhui 230026, China}
\affiliation{Peng Huanwu Center for Fundamental Theory, Hefei, Anhui 230026, China}
%\date{\today.}

\begin{abstract}
In this paper we consider a parity violating gravity model without higher derivatives, thus ghost free. This model is constructed from the tetrad and its derivatives and coupled to a dynamical scalar field, like axion. It can be reduced from the Nieh-Yan term within the framework of teleparallel gravity. We apply this model to cosmology and investigate its consequences on cosmological perturbation theory. We find that the coupled dynamical scalar field lost its independent dynamics at the linear order and the parity violating term itself behaves like a viscosity. For gravitational waves, this model produces velocity difference between left- and right-handed polarizations, but generates no amplitude discrepancy.
\end{abstract}

\maketitle

\section{introduction}

Discrete symmetries, including the charge conjugation (C), parity (P), time reversal (T) and their combinations have played important roles in discovering fundamental physical laws. Since the parity violation was discovered in weak interactions \cite{Lee:1956qn},  we gradually know that most of them are not exact in the nature. Whether or not these discrete symmetries are broken in gravity is unclear to us, but recently there were lots of studies on the parity violating (PV) gravities in the literature, stimulated by the experimental detections of gravitational waves (GWs) \cite{Abbott:2016blz,TheLIGOScientific:2017qsa} and the developments in the cosmic microwave background radiation (CMB) experiments, aiming to find primordial GWs originated from the early universe. If there are parity violations in gravity, the left and right polarized GWs should have different behaviors and corresponding signals are possibly to be captured by well-designed experiments.

One well-studied PV gravity is the Chern-Simons (CS) gravity \cite{Jackiw:2003pm,Alexander:2009tp}, which modifies general relativity (GR) through introducing a gravitational CS term (in addition to the Einstein-Hilbert term) in the action: $S_{CS}\sim \int d^4x \sqrt{-g} \theta(x) \varepsilon^{\mu\nu\rho\sigma}R_{\mu\nu}^{~~\alpha\beta}R_{\rho\sigma\alpha\beta}$, here $g$ is the determinant of the metric $g_{\mu\nu}$, $\theta(x)$ is the coupling scalar field,  $\varepsilon^{\mu\nu\rho\sigma}$ is the four dimensional Levi-Civita tensor defined in terms of the antisymmetric symbol $\epsilon^{\mu\nu\rho\sigma}$ as $\varepsilon^{\mu\nu\rho\sigma}=\epsilon^{\mu\nu\rho\sigma}/\sqrt{-g}$, and $R_{\rho\sigma\alpha\beta}$ is the Riemann tensor constructed from the metric. At the linear perturbation level, the CS modification makes a difference between the amplitudes of the left- and right-handed polarizations of GWs, but their velocities remain the same. This phenomenon sometimes is called amplitude birefringence. Even though the CS term is the leading order PV modification to GR from the viewpoint of metric theory, it contains the product of two Riemann tensors and leads to a higher-derivative field equation, because each Riemann tensor hides a second derivative of the metric. Higher derivative equation often implies the existence of ghost mode. This is indeed the case in the CS gravity, as demonstrated in \cite{Dyda:2012rj}, one polarized component of GWs becomes ghost at the region with large wave number and causes vacuum instability.

To avoid ghost, further extensions to the CS gravity has been explored in \cite{Crisostomi:2017ugk} where several terms including the first or higher derivatives of the coupling scalar field $\theta(x)$ were introduced into the action. Though each term contains higher derivatives (of the metric or the coupling field), special combinations of them can be elaborated to prevent the higher than second order time derivatives from appearing in the equations of motion, thus eliminating the ghost modes. Such kind of models with more complex forms were studied recently in \cite{Gao:2019liu}. A common feature of this kind of PV gravity models is that besides the amplitude birefringence, the discrepancy between the phase velocities of GWs with different handedness (called velocity birefringence) is also produced. Some other PV gravity models from different motivations may be found, e.g., in the references of \cite{Zhao:2019xmm}. All of them have forms much more complex than the CS gravity.

The question we ask in this paper is whether we can find a simple and healthy PV gravities without introducing higher derivatives. From the view of metric theory based on Riemannian geometry, no PV term simpler than CS can be constructed from the Riemann and Levi-Civita tensors for the action. However, there are other approaches to gravity theories beyond the metric formulation. Examples which were well-studied in recent years are the teleparallel gravity (TG) \cite{Tele} and the symmetric teleparallel gravity (STG) models \cite{Nester:1998mp}. Both models are equivalent to GR but formulated with flat spacetime, i.e., the curvature vanishes. In TG model, gravitation is attributed to torsion which contains up to the first derivatives of the tetrad fields, nevertheless in STG model gravitation is identified with the non-metricity which contains only the first derivatives of the metric. These formulations provide flexible frameworks for generalizations, especially for the purpose of constructing PV gravity models without higher derivatives. In fact, PV extensions of the theory equivalent to GR has been studied recently within the framework of STG model \cite{Conroy:2019ibo}. In this paper, we will realize this idea in the framework of TG. We will propose a PV gravity model which is healthy and simple in form, and then study its implications in cosmology.

\section{GR equivalent Teleparallel gravity}

First, we will make a brief introduction of TG theory. The building blocks of TG theory are the tetrad fields (or vielbeins), $e^A_{~\mu}$, which were used to construct the spacetime metric $g_{\mu\nu}=\eta_{AB}e^A_{~\mu}e^B_{~\nu}$ with $\eta_{AB}={\rm diag}\{+1,-1,-1,-1\}$ being the metric of local (Minkowskian) space. From now on, we will use the unit $8\pi G=1/M_p^2=1$, and $A, B, C,...=0, 1, 2, 3$ and $a, b, c,...=1, 2, 3$ to denote the tetrad indices. The spacetime tensor indices are represented by $\mu, \nu, \rho,...=0, 1, 2, 3$ and $i, j, k,...=1, 2, 3$. The tetrad indices will be lowered and risen by $\eta_{AB}$ and $\eta^{AB}$, and the spacetime indices by the metric $g_{\mu\nu}$ and its inverse. The antisymmetric symbol $\epsilon^{\mu\nu\rho\sigma}$ has the properties: $\epsilon^{0ijk}=\epsilon^{ijk}=-\epsilon_{ijk}$, and $\epsilon^{123}=1$. 
TG theory starts from a flat spacetime where the curvature two form constructing from the spin connection vanishes:
\be
\hat{R}^A_{~B\mu\nu}=\partial_{\mu}\omega^A_{~B\nu}+\omega^A_{~C\mu}\omega^C_{~B\nu}-\{\mu\leftrightarrow\nu\}=0~,
\ee
 so does the Riemann tensor from the spacetime connection:
 \be
 \hat{R}^{\rho}_{~\sigma\mu\nu}=e_A^{~\rho}e^B_{~\sigma}\hat{R}^A_{~B\mu\nu}=\partial_{\mu}\hat{\Gamma}^{\rho}_{\nu\sigma}+\hat{\Gamma}^{\rho}_{\mu\alpha}\hat{\Gamma}^{\alpha}_{\nu\sigma}-\{\mu\leftrightarrow \nu\}=0~.
 \ee
 This puts a constraint on the spin connection so that it generally has the form $\omega^A_{~B\nu}=(\Lambda^{-1})^A_C\partial_{\nu}\Lambda^C_B$ with $\Lambda^C_B$ being the matrix elements of Lorentz transformation, considering as functions of spacetime. In addition, the metricity demands that $\omega_{AB\nu} =-\omega_{BA\nu}$.  The gravity is identified with the torsion
\be
\mathcal{T}^{\rho}_{~\mu\nu}=\hat{\Gamma}^{\rho}_{\mu\nu}-\hat{\Gamma}^{\rho}_{\nu\mu}~.
\ee
The GR equivalent TG theory has the action
\bea\label{tgaction}
S_g=\frac{1}{2}\int d^4x ~{\rm e}\mathbb{T}=\int d^4x~{\rm e} (-\frac{1}{2}\mathcal{T}_{\mu}\mathcal{T}^{\mu}+\frac{1}{8}\mathcal{T}_{\alpha\beta\mu}\mathcal{T}^{\alpha\beta\mu}+\frac{1}{4}\mathcal{T}_{\alpha\beta\mu}\mathcal{T}^{\beta\alpha\mu})~,
\eea
where ${\rm e}=\sqrt{-g}$ is the determinant of the tetrad $e^A_{~\mu}$, $\mathbb{T}$ is the torsion scalar, and $\mathcal{T}_{\mu}=\mathcal{T}^{\alpha}_{\ \mu\alpha}$ is the torsion vector. This action is identical to the Einstein-Hilbert action of GR up to a surface term
\be\label{graction}
S_g=\int d^4x \sqrt{-g}[-\frac{1}{2} R (e)-\nabla_{\mu}\mathcal{T}^{\mu}]~,
\ee
the curvature scalar $R(e)$ and the covariant derivative $\nabla_{\mu}$ are associated with the Levi-Civita connection, i.e., the Christoffel symbol. Giving up the surface term, the equivalent action (\ref{graction}) is fully constructed from the tetrad fields and their derivatives. The spin connection is pure gauge in the TG action (\ref{tgaction}). Due to this fact, the Weitzenb\"{o}ck connection with $\omega^A_{~B\nu}=0$, was usually taken in the literature, with this gauge choice the original spacetime connection is simply,
\be\label{gauge}
\hat{\Gamma}^{\rho}_{\mu\nu}=e_A^{~\rho}\partial_{\mu}e^A_{~\nu}~,
\ee
so that the torsion and the torsion two form are simply
\bea\label{simple}
\mathcal{T}^{\rho}_{\ \mu\nu}=e_A^{~\rho}(\partial_{\mu}e^A_{~\nu}-\partial_{\nu}e^A_{~\mu})~,~\mathcal{T}^A_{~\mu\nu}=e^A_{~\rho}\mathcal{T}^{\rho}_{~\mu\nu}=\partial_{\mu}e^A_{~\nu}-\partial_{\nu}e^A_{~\mu}~.
\eea
The most-studied generalizations of TG theory are the so-called $f(\mathbb{T})$ models \cite{Bengochea:2008gz}, in which the torsion scalar in the action (\ref{tgaction}) is replaced by its arbitrary functions. They were found \cite{Yang:2010ji} to be different from $f(R)$ generalizations to GR, though $\mathbb{T}$ is equivalent to $R$.  See the review \cite{Cai:2015emx} and references therein for the applications to cosmology. Recently there were also some Horndeski type generalizations of TG theory, see e.g.,  \cite{Bahamonde:2019shr,Bahamonde:2019ipm,Bahamonde:2020cfv}.

\section{Parity violating extension from the view point of TG}

In this paper we propose a PV gravity model by introducing additional PV terms into the GR equivalent TG action (\ref{tgaction}), so this is equivalently making a PV extension to GR. In terms of the torsion language, it is natural to consider the scalar field coupled Nieh-Yan term \cite{Nieh:1981ww, Nieh:2013ada},
\be
S_{NY}=\int d^4x ~{\rm e} ~\frac{c \theta}{4} (\mathcal{T}_{A\mu\nu}\widetilde{\mathcal{T}}^{A\mu\nu}-\varepsilon^{\mu\nu\rho\sigma}\hat{R}_{\mu\nu\rho\sigma})~,
\ee
where $c$ is the coupling constant and $\widetilde{\mathcal{T}}^{A\mu\nu}=(1/2)\varepsilon^{\mu\nu\rho\sigma}\mathcal{T}^A_{~~\rho\sigma}$ is the dual of the torsion two form $\mathcal{T}^A_{~~\mu\nu}$. 
The Nieh-Yan term ${\rm e}~ (\mathcal{T}_{A\mu\nu}\widetilde{\mathcal{T}}^{A\mu\nu}-\varepsilon^{\mu\nu\rho\sigma}\hat{R}_{\mu\nu\rho\sigma})$ itself is a topological density and was studied extensively in Riemann-Cartan theories. If $\theta$ is a constant, the Nieh-Yan term only contributes a surface term and will have no effect on the equations of motion. 
With position dependent $\theta$, this action breaks the parity symmetry of the gravitational field if $\theta$ has a non-vanishing background\footnote{Generally CP and CPT symmetries are also violated. CP violation is due to the charge blind feature of the graviton. CPT is violated due to T conservation in a static background or T breaking in an evolving background, in the latter case T violation in general cannot compensate the CP violation.}. 
Theoretically, the $\theta$ field can be thought as axion-like, because the Nieh-Yan action is invariant under the shift $\theta\rightarrow \theta+\theta_0$ by a constant $\theta_0$. Such a coupling can appear in the mechanisms \cite{Mercuri:2009zi, Castillo-Felisola:2015ema}  to regularize the infinities in theories of the Einstein-Cartan manifold, similar to the QCD axion coupling in the Peccei-Quinn mechanism \cite{Peccei:1977hh} for a solution to the strong CP problem. 

In stead of considering Einstein-Cartan, in this paper we start from the TG model. In the spirit of teleparallelism, the second term associated with curvature in the Nieh-Yan action drops out. In addition, we also take the Weitzenb\"{o}ck connection, as usually done in studies of $f(\mathbb{T})$ models. With these considerations, the Nieh-Yan action reduces to
\be\label{coupling}
S_{NY}=\int d^4x \sqrt{-g} ~\frac{c \theta}{4}\mathcal{T}_{A\mu\nu}\widetilde{\mathcal{T}}^{A\mu\nu}=\int d^4x \sqrt{-g} ~\frac{c \theta}{8}\eta_{AB}\varepsilon^{\mu\nu\rho\sigma}(\partial_{\mu}e^A_{~\nu}-\partial_{\nu}e^A_{~\mu})(\partial_{\rho}e^B_{~\sigma}-\partial_{\sigma}e^B_{~\rho})~.
\ee
This is the coupling what we actually introduced. In above we showed the motivation to it by introducing the Nieh-Yan term to the GR equivalent TG theory (again, not Einstein-Cartan). In fact there is another motivation, that is to consider the coupling (\ref{coupling}) directly in GR where the tetrad fields are treated as fundamental. From this point, the gravity is considered as a theory about four non-vanishing vector fields, $e^A_{\ \mu}$, and Eq. (\ref{coupling}) is the anomalous coupling between these four vector fields and a scalar field, very similar to the CS coupling in electrodynamics \cite{Carroll:1989vb}. This coupling breaks the local Lorentz symmetry, but the diffeomorphism is preserved. It is not invariant under position-dependent Lorentz transformation: $e^A_{\ \mu}\rightarrow \Lambda^A_B(x) e^B_{\ \mu}$. This does not contradict current experiments. Similar to what happens in $f(\mathbb{T})$ models, this may be considered as the consequence led by a specific gauge choice, the Weitzenb\"{o}ck connection, from the viewpoint of TG theory.

We also take into account the kinetic and potential terms of the scalar field and the action of other matter which coupled minimally through the metric (or the tetrad). The full action in its equivalent form is
\bea\label{model}
S=S_g+S_{NY}+S_{\theta}+S_m=\int d^4x \sqrt{-g}[-\frac{R}{2}+\frac{c \theta}{4}\mathcal{T}_{A\mu\nu}\widetilde{\mathcal{T}}^{A\mu\nu}+\frac{1}{2}\nabla_{\mu}\theta\nabla^{\mu}\theta-V(\theta)]+S_m~,
\eea
where $S_g$ is the same as that in Eq. (\ref{tgaction}), but at the second step we have neglected the surface term (see Eq. (\ref{graction})), which has no effect on the equations of motion. The torsion $\mathcal{T}^A_{~~\mu\nu}$ is given by Eq. (\ref{simple}). The gravitational field equation follows from the variational principle while treating $e^A_{~\mu}$ as basic variable,
\be
(G^{\mu\nu}-T^{\mu\nu}-T^{\mu\nu}_{\theta})e^A_{~\nu}+c \partial_{\nu}\theta\widetilde{\mathcal{T}}^{A\mu\nu}=0~,
\ee
where $G^{\mu\nu}$ is the Einstein tensor, $T^{\mu\nu}=-(2/\sqrt{-g})(\delta S_m/\delta g_{\mu\nu})$ and
$T^{\mu\nu}_{\theta}=[V(\theta)-\nabla_{\alpha}\theta\nabla^{\alpha}\theta/2]g^{\mu\nu}+\nabla^{\mu}\theta\nabla^{\nu}\theta$ are the energy-momentum tensors for the matter and the scalar field respectively.
The tetrad fields $e^A_{~\nu}$ are not degenerate, we may multiply their inverses to above equation, so the equation of motion can be rewritten as,
\be\label{eom}
G^{\mu\nu}+N^{\mu\nu}=T^{\mu\nu}+T^{\mu\nu}_{\theta}~,
\ee
where the tensor $N^{\mu\nu}=c e_A^{~\nu}\partial_{\rho}\theta\widetilde{\mathcal{T}}^{A\mu\rho}=c \partial_{\rho}\theta\widetilde{\mathcal{T}}^{\nu\mu\rho}$. The equation (\ref{eom}) contains $16$ instead of $10$ equations, because the basic variables are the tetrad fields which have totally $16$ components. Among them, $10$ equations corresponding to the components symmetric under the permutation of $\mu$ and $\nu$ are the Einstein field equations modified by $N^{(\mu\nu)}\equiv(1/2)(N^{\mu\nu}+N^{\nu\mu})$, here the parentheses represent the symmetrization of indices. The rest $6$ equations state that the antisymmetric part of $N^{\mu\nu}$ vanish, i.e., $N^{[\mu\nu]}\equiv(1/2)(N^{\mu\nu}-N^{\nu\mu})=0$. This means that $N^{\mu\nu}$ is constrained to be symmetric, 
\be\label{symconstraint}
N^{\mu\nu}=N^{\nu\mu}~. 
\ee
As we will show later, this is a strong constraint on the coupled system.
In addition, the Bianchi identity and the covariant conservation law, $\nabla_{\mu}T^{\mu\nu}=0$, demand that $\nabla_{\mu}N^{\mu\nu}=\nabla_{\mu}T^{\mu\nu}_{\theta}$. This brings no further constraint. In fact one can prove $\nabla_{\mu}N^{\mu\nu}=(c/4) \mathcal{T}_{A\rho\sigma}\widetilde{\mathcal{T}}^{A\rho\sigma}\nabla^{\nu}\theta$ and $\nabla_{\mu}T^{\mu\nu}_{\theta}=(\Box\theta+V_{\theta})\nabla^{\nu}\theta$, here and in the following we will use the notations $V_{\theta}$ and $V_{\theta\theta}$ to denote the first and second derivatives of the potential to the scalar field. The requirement $\nabla_{\mu}N^{\mu\nu}=\nabla_{\mu}T^{\mu\nu}_{\theta}$ is consistent with the Klein-Gordon equation:
\be
\Box\theta+V_{\theta}-\frac{c}{4} \mathcal{T}_{A\mu\nu}\widetilde{\mathcal{T}}^{A\mu\nu}=0~,
\ee
which is obtained from the variation of the action (\ref{model}) with respect to the scalar field $\theta$. 

We would like to comment here why we should treat $\theta$ as a dynamical field by including its kinetic and potential terms in the action. If $\theta$ is just a non-dynamical parameter, the tensor $N^{\mu\nu}$ would be constrained to be divergenceless, i.e., $\nabla_{\mu}N^{\mu\nu}=0$, this in turn requires a vanishing Pontryagin density for the tetrad fields: $\mathcal{T}_{A\rho\sigma}\widetilde{\mathcal{T}}^{A\rho\sigma}=0$ . This extra constraint together with the constraint from the permutation symmetry (\ref{symconstraint}) put strong restrictions on the space of solutions. One will find that only few solutions to the gravitational field equation, e.g., the Schwarzschild solution and the spatially flat Friedmann-Robertson-Walker (FRW) solution, can be easily obtained. Some other solutions common in GR, like Kerr and spatially curved FRW solutions, are hard to found if not impossible. This is similar to the case of non-dynamical CS gravity \cite{Jackiw:2003pm,Alexander:2009tp}. Furthermore, in this case the equations are not closed. There is no equation of motion for $\theta$, it can only be determined by hand. 
If $\theta$ is a dynamical field, these will not happen. For instance, Kerr solution can be satisfied with some configurations of $\theta$, these configurations themselves are solutions to the Klein-Gordon equation subject to some boundary conditions. As mentioned above, in our case, the Bianchi identity brings no further constraint and the Nieh-Yan action just describes a non-minimal coupling between the scalar field and the gravitational field.

\section{Application to cosmology}

Now we apply our PV gravity model (\ref{model}) to cosmology. First we consider the background: a spatially flat FRW universe, where the tetrad fields are parametrized as $e^A_{~\mu}=a(\eta)\delta^A_{\mu}$ so that the line element for the spacetime is $ds^2=a^2(d\eta^2-\delta_{ij}dx^idx^j)$. Here $a(\eta)$ is the scale factor of the universe and $\eta$ is the conformal time. The equations are the same as those of GR with minimally coupled $\theta$ field and matter
\be\label{ba}
3\mathcal{H}^2=a^2(\rho_{\theta}+\rho)~,~2\mathcal{H}'+\mathcal{H}^2=-a^2(p_{\theta}+p)~,~\theta''+2\mathcal{H}\theta'+a^2V_{\theta}=0~,
\ee
where prime represents the derivative with respect to the conformal time, $\mathcal{H}=a'/a=aH$ is the conformal Hubble rate, $\rho_{\theta}=\theta'^2/(2a^2)+V$ and $p_{\theta}=\theta'^2/(2a^2)-V$ are the energy density and pressure of the $\theta$ field, and $\rho$ and $p$ denote the energy density and pressure of other matter. All of them only depend on time.
This means the Nieh-Yan coupling has no effect in the spatially flat FRW universe, consistent with what we have commented above on the solutions of the model with non-dynamical $\theta$ parameter.

To explore the modifications brought by this model, we turn to the linear cosmological perturbation theory. We use the parametrization for the tetrad fields proposed in Refs. \cite{Izumi:2012qj,Golovnev:2018wbh}:
\bea
& & e^{0}_{\ 0}=a(1+A)~,~\nonumber e^{0}_{\ i}=a(\partial_{i}\beta+\beta_{i}^{V})~,~\nonumber e^{a}_{\ 0}=a\delta_{ai}(\partial_i\gamma+\gamma_{i}^{V})~,\nonumber\\
& & e^{a}_{\ i}=a\delta_{aj}[ (1-\psi)\delta_{ij}+\partial_{j}\partial_{i}\alpha+\partial_{i}\alpha_{j}^{V}+
              \epsilon_{ijk}(\partial_{k}\lambda+\lambda_{k}^{V})+\frac{1}{2}h^{T}_{ij}]~,
\eea
so that the perturbed metric components have the following forms:
\bea
& &g_{00}=a^{2}(1+2A)~,~ g_{0i}=-a^{2}(\partial_{i}(\gamma-\beta)+\gamma_{i}^{V}-\beta_{i}^{V})~,\nonumber\\
& &g_{ij}=-a^{2}[(1-2\psi)\delta_{ij}+2\partial_{i}\partial_{j}\alpha+\partial_{i}\alpha_{j}^{V}+\partial_{j}\alpha_{i}^{V}+h^{T}_{ij}]~.
\eea
Besides the scalar perturbations: $A, \gamma-\beta, \psi, \alpha$, vector perturbations: $\gamma_i^V-\beta_i^V, \alpha^V_i$, and tensor perturbation: $h^T_{ij}$ in the metric, the parametrization of tetrad brings extra scalar perturbation $\lambda$ and vector perturbation $\lambda_i^V$. All the vector perturbations are transverse and denoted by the superscript $V$, all the tensor perturbations are transverse and traceless and denoted by the superscript $T$. The $\theta$ field is decomposed as $\theta(\eta, \vec{x})=\theta(\eta)+\delta\theta$. Other matter, considered as fluid, has the following energy-momentum tensor up to the linear order:
\bea
& & T^{0}_{\ 0}=\rho+\delta\rho~,~T^{0}_{\ i}=(\rho+p)(\partial_{i}v+v_{i}^{V})\nonumber\\
& & T^{i}_{\ j}=-(p+\delta p)\delta^i_{j}+\Sigma^i_{\ j}~.
\eea
The velocity perturbation contains scalar and vector perturbations. The anisotropic stress $\Sigma^i_{\ j}=\partial_{i}\partial_{j}\sigma-(\frac{1}{3}\nabla^{2}\sigma)\delta_{ij} +\partial_{(i}\sigma_{j)}^{V}+\sigma^{T}_{ij}$
is traceless, it can also be generally decomposed into the scalar, vector and tensor perturbations.

Linear perturbation equations are obtained by substituting above parametrizations to Eqs. (\ref{eom}) and (\ref{symconstraint}).
We will transform to the Fourier space, so each $\partial_i$ is replaced by $ik_i$.
First, the equation (\ref{symconstraint}) gives the constraint
\be\label{surprise}
\theta' \psi+\mathcal{H}\delta\theta=0~,
\ee
on scalar perturbations, and
\be
~\beta_i^V=\lambda_i^V=0~,
\ee
on vector perturbations. It puts no further constraint on tensor perturbations.
These constraints are gauge-invariant. The constraint on vector perturbations brings no important modification. But the result (\ref{surprise}) on scalar perturbations is surprising, it means the curvature perturbation of the hypersurface with homogeneous $\theta$, denoted by $-\psi-\mathcal{H}\delta\theta/\theta'$, vanishes identically. At the beginning, the $\theta$ field was introduced as an independent dynamical field, but here we see that when it couples to the tetrad field through the Nieh-Yan term, it cannot fluctuate independently, at least at the linear order. Whether this dynamical degree of freedom is excited at higher orders deserves further investigations.

Then, we turn to the perturbative gravitational field equations from Eq. (\ref{eom}).
As shown there, it is the tensor $N^{\mu}_{\ \nu}$ that describes the deviations from the Einstein field equations of GR.  This tensor has only the following non-zero components at the linear order:
\be
N^i_{\ j}=\frac{c\theta'}{a^2}(\partial_i\partial_j\lambda-\nabla^2\lambda \delta_{ij})+\frac{c\theta'}{2a^2}\epsilon^{ikl}\partial_kh^T_{jl}~.
\ee
These components do not contain any vector perturbation. So we can expect that the equations for vector perturbations from Eq. (\ref{eom}) are not changed, and we will neglect vector perturbations at the rest of this paper. We can also expect that the model modify the scalar perturbation equations through the variable $\lambda$, which is only a tetrad perturbation, not presented in the metric. It was named as pseudo scalar perturbation in Refs. \cite{Izumi:2012qj,Golovnev:2018wbh}.  The full set of scalar perturbation equations from Eq. (\ref{eom}) in the conformal Newtonian gauge $\gamma-\beta=0, \alpha=0$ are listed below:
\bea\label{per}
 && 2k^{2}\psi+6\mathcal{H}(\psi'+\mathcal{H}A)=-(\theta'\delta\theta'- \theta'^{2}A+a^{2}V_{\theta}\delta\theta)-a^{2}\delta\rho\nonumber\\
&& 2\psi'+2\mathcal{H}A= \theta'\delta\theta+ a^{2} (\rho+p)v\nonumber\\
 &&2\psi''+2\mathcal{H}(A'+2\psi')+(2\mathcal{H}^{2}+4\mathcal{H}')A= (\theta' \delta\theta'- \theta'^{2}A-a^{2}V_{\theta}\delta\theta)+a^{2}(\delta p+\frac{2}{3}k^{2}\sigma)\nonumber\\
 &&\psi-A=c\theta'\lambda-a^2 \sigma~.
\eea
The modification to GR by our model appeared in the last equation. It shows that $\lambda$ behaves like a viscosity, imprints the imperfect fluid nature and makes a difference between the perturbations $\psi$ and $A$. To calculate this viscosity, we need also to take into account the constraint (\ref{surprise}) and
the perturbed Kein-Gordon equation
\begin{equation}\label{theta2}
  \delta\theta''+2\mathcal{H}\delta\theta'+k^{2}\delta\theta+a^{2}V_{\theta\theta}\delta\theta-\theta'(A'+3\psi')+2a^{2}V_{\theta}A=-2c\mathcal{H}k^2\lambda~.
\end{equation}

The equation for tensor perturbations from Eq. (\ref{eom}) is
\be\label{tensoreom}
{h^{T}_{ij}}''+2\mathcal{H}{h^{T}_{ij}}'+k^{2}h^{T}_{ij}+c \theta' (i k_{l})\epsilon_{lk(i} h^{T}_{j)k}=-2 a^{2} \sigma^{T}_{ik}~,
\ee
the last term at the left hand side indicates parity symmetry breaking. This is more obvious if we turn to expanding the tensor perturbations in the circular polarization bases $\hat{e}^{L}_{ij}$ and  $\hat{e}^{R}_{ij}$,
\bea
& & \nonumber h^{T}_{ij}=h^{L}\hat{e}^{L}_{ij}+h^{R}\hat{e}^{R}_{ij}\\
& & \nonumber \sigma^{T}_{ij}=\sigma^{L}\hat{e}^{L}_{ij}+\sigma^{R}\hat{e}^{R}_{ij}~.
\eea
The bases satisfy the relation: $n_{l}\epsilon_{lik}\hat{e}^{A}_{jk}=i\lambda_{A}\hat{e}^{A}_{ij}$ , here $A=L, R$ and $\lambda_{L}=-1, \lambda_{R}=1$, $\vec{n}$ is the unit vector of $\vec{k}$.
So the equation (\ref{tensoreom}) can be rewritten as
\be\label{tensoreom1}
{h^{A}}''+2\mathcal{H}{h^{A}}'+(k^2+c\lambda_{A} \theta'k)h^{A}=-2 a^{2} \sigma^{A}~.
\ee
Without the source this equation shows that left- and right-handed polarized GWs propagate with different velocities. To make more clear, the sourceless equation (\ref{tensoreom1}) is further rewritten as ${v^{A}}''+(\omega_A^2-a''/a)v^{A}=0$ after a renormalization: $v^A=ah^A$.  Where $\omega_A^2=k^2+\lambda_A c\theta'k=k^2(1+\lambda_A c\theta'/k)\equiv k^2(1+\mu_A)$ is the modified dispersion relation. The dispersion relation shows how the waves with various frequencies propagate within the horizon, where $a''/a$ can be neglected. In terms of the notation of Ref. \cite{Qiao:2019wsh}, the deviation from the standard dispersion relation is characterized by the parameter $\mu_A$. Consider small coupling and slow evolution of $\theta$, one can find that GWs with different helicities will have different phase velocities: $v^A_p=\omega_A/k\simeq 1+\lambda_A c\theta'/(2k)$, and same group velocity $v_g=d\omega_A/dk\simeq 1+c^2\theta'^2/(8k^2)$ up to the order $\mathcal{O}(c^2)$. This is the so-called velocity birefringence phenomenon of GWs, very similar to the cosmic birefringence induced by electromagnetic Chern-Simons coupling \cite{Carroll:1998zi,Lue:1998mq,Feng:2006dp,Li:2008tma}. Since the deviation $\mu_A$ is inversely proportional to $k$, this is an infrared effect, contrary to most PV gravity models in the literature. We can also see that the phase velocity difference or the deviation of the group velocity from the speed of light in vacuum become important only at the region of small $k$ (large scales). The amplitudes for both helicities are the same, this is different from the CS gravity. Another difference from CS gravity is that in this model there is no higher derivative in the equation of motion for tensor perturbations. We note that the velocity birefringence phenomenon similar to ours can also be obtained in the PV gravity model from STG theory \cite{Conroy:2019ibo}.

We would like to point out again the difference between the cases with dynamical and non-dynamical $\theta$ fields. If $\theta$ is non-dynamical, it does not represent a species in the universe. The energy density $\rho_{\theta}$ and the pressure $p_{\theta}$ drop out from the first and second equations of Eq. (\ref{ba}), and the third equation is absent. At the linear level, Eq. (\ref{theta2}) reduces to $\lambda=0$. So all the terms related to $\theta$ and $\lambda$ are absent from the scalar perturbation equations (\ref{per}). All these are the same with the equations of the universe filled by fluids excluding $\theta$ within the context of GR. The Nieh-Yan term only makes a difference in the tensor perturbation equation (\ref{tensoreom}), but there is no equation to determine $\theta'$, it can only be predetermined by hand.

\section{quadratic actions for scalar and tensor perturbations}

When applying this PV gravity to the early universe, such as the inflationary epoch, we should care about the problem of quantum originated primordial perturbations. For this purpose, we need quadratic actions for the scalar and tensor perturbations. Because the curvature perturbation corresponding to $\theta$ vanishes, see Eq. (\ref{surprise}), it cannot be used to generate the density perturbation. We simply consider the model in Eq. (\ref{model}) where the matter (described by $S_m$) is just another minimally coupled scalar field, i.e., $S=S_g+S_{NY}+S_{\theta}+S_{\phi}$
with
\be
S_{\phi}=\int d^4x\sqrt{-g}[\frac{1}{2}\nabla_{\mu}\phi\nabla^{\mu}\phi-U(\phi)]~.
\ee
The curvature perturbation of hypersurfaces with homogeneous $\phi$ field is denoted by $\zeta=-(\psi+\mathcal{H}\delta\phi/\phi')$. For our purpose, it is vey convenient  to choose the unitary gauge where $\delta\phi=0~,~\alpha=0$, so that $\zeta=-\psi$. After some tedious calculations, the quadratic action for scalar perturbations is
\bea\label{scalar2}
\nonumber   S^{(2)}&=& \int d^{4}x \, a^{2}\bigg\{-\Big[2A\zeta_{,ii}-\zeta_{,i}\zeta_{,i}+3\mathcal{H}^{2}(6\zeta^{2}+A^{2})\\
\nonumber &+&2\mathcal{H}(9\zeta\zeta'-3\zeta'A+A(\gamma-\beta)_{,ii})+3\zeta'^{2}-2\zeta'(\gamma-\beta)_{,ii}+(\mathcal{H}'-\mathcal{H}^{2})A^{2}+9\mathcal{H}'\zeta^{2}\Big]\\
\nonumber &+&\left[\frac{1}{2}\delta\theta'^{2}-\frac{1}{2}\delta\theta_{,i}\delta\theta_{,i}-\frac{1}{2}a^{2}V_{\theta\theta}\delta\theta^{2}
           +\theta'(A'-3\zeta')\delta\theta-2a^{2}V_{\theta}A\delta\theta+\theta'\delta\theta(\gamma-\beta)_{,ii}\right]\\
\label{scalaraction1} &+&2c (\mathcal{H}\delta\theta-\theta'\zeta)\lambda_{,ii}\bigg\}~.
\eea
The variables $\lambda$ and $\gamma-\beta$ are Lagrange multipliers, and $A$ is also an auxiliary variable. All these three non-dynamical fields induce the following constraints:
\bea
& & \mathcal{H}\delta\theta-\theta'\zeta=0~,\\
& & 2\mathcal{H}A-2\zeta'- \theta'\delta\theta=0~,\\
& & 2\mathcal{H}(\gamma-\beta)_{,ii}+2(\mathcal{H}'+2\mathcal{H}^{2})A-6\mathcal{H}\zeta'+2\zeta_{,ii}+(\theta'\delta\theta'+V_{\theta}\delta\theta)=0~.
\eea
The first equation showed again that $\delta\theta$ is not an independent perturbation. Substituting these constraints back into the action (\ref{scalar2}), we get the final form of the quadratic action for scalar perturbation:
\begin{equation}\label{scalaraction2}
  S^{(2)}=\int d^{4}x\; z^{2}\left(\frac{1}{2}\zeta'^{2}-\frac{1}{2}\zeta_{,i}\zeta_{,i}-\frac{1}{2}\bar{m}^{2}\zeta^{2}\right)
\end{equation}
where $z^2=a^{2}\phi'^2/\mathcal{H}^2$, and the effective mass square
\be
 \bar{m}^{2}=-\frac{\theta'^2}{\mathcal{H}^{2}}[2\mathcal{H}^2+\mathcal{H}'+
a^2\mathcal{H}(\frac{U_{\phi}}{\phi'}+\frac{V_{\theta}}{\theta'})]=~-\frac{a^2\theta'^2}{\mathcal{H}^{2}}[U(\phi)+V(\theta)+
\mathcal{H}(\frac{U_{\phi}}{\phi'}+\frac{V_{\theta}}{\theta'})].
\ee
Again, this quadratic action showed clearly that there is only one dynamical scalar degree of freedom, even though we introduced two scalar fields at the beginning. 
In another word, the field $\theta$ does not provide an entropy (or isocurvature) perturbation. This quadratic action (\ref{scalaraction2}) is very different from those of traditional single field or double fields inflation models (or models alternative to inflation). This may be useful for searching new mechanism of generating primordial perturbations. We left this consideration in the future works.

It is straightforwardly to obtain the quadratic action for the tensor perturbations
\begin{equation}\label{tensoraction}
  S=\int d^{4}x \ \frac{a^{2}}{8} \left({h^{T}_{ij}}'{h^{T}_{ij}}'-\partial_{k}h^{T}_{ij}\partial_{k}h^{T}_{ij}
  -c\theta' \epsilon_{ijk}h^{T}_{il}\partial_{j}h^{T}_{kl}\right)~.
\end{equation}
After expanding the tensor perturbation in Fourier space by circular polarization bases
\begin{equation}
  h^{T}_{ij}(t,\vec{x})=\sum_{A=L,R}\int \frac{d^{3}k}{(2\pi)^{3/2}}\ h^{A}(t,\vec{k})\ \hat{e}^{A}_{ij}(\vec{k})\ e^{ik_{j}x^{j}}~,
\end{equation}
the quadratic action can be rewritten as
\begin{equation}\label{tensoraction2}
  S=\sum_{A=L,R}\int d\eta d^{3}k \ \frac{a^{2}}{4} \left[ {h^{A*}}'{h^{A}}'-(k^{2}+c \theta'\lambda_{A}k)h^{A*}h^{A}\right]~.
\end{equation}
These quadratic actions showed clearly that there is no ghost instability in this model, contrary to the CS gravity. Again, it only produces velocity birefringence rather than amplitude birefringence phenomenon of GWs.
It is expectable that from the quadratic action (\ref{tensoraction}), the produced primordial gravitational waves in the early universe will have different power spectra for left- and right-handed polarizations. This will result in the correlation between the E- and B-modes polarizations of the cosmic microwave background radiation (CMB), and may have effects on the planned CMB experiments \cite{Li:2017drr, Abazajian:2019eic}. 

\section{Conslusions}

We showed in this paper that a simple and healthy parity violating gravity can be obtained. This is a small extension of general relativity. The parity violating term is introduced from the view point of teleparallel gravity in terms of the tetrad and torsion language, and can be considered as reduced from the Nieh-Yan term coupling with a dynamical scalar field. We also applied this model to cosmology and found that the coupled scalar field does not have independent dynamics at the linear perturbation level, though it is indeed dynamical at the background. The tensor perturbations are ghost free and present the velocity birefringence phenomenon. No amplitude birefringence produced in this model. In the future, we will study the primordial perturbations generations within this framework in terms of some specific scalar field models. In addition, the question why the coupled scalar field cannot have independent dynamics deserves further studies.

{\it Acknowledgement}: This work is supported by NSFC under Grant No. 12075231, 11653002, and 11947301.

{\it Note added}: While this manuscript is preparing for submission, there appeared in eprint arXiv a paper \cite{Chatzistavrakidis:2020wum} in which the authors proposed a same parity violating term but coupled to a non-dynamical $\theta$ field and studied its consequences in gravitoelectromagnetism.

{}

\end{document}